\documentclass[showpacs,preprintnumbers,amsmath,amssymb,aps, prd]{revtex4}
\usepackage{graphicx}
\usepackage{grffile}
\usepackage{caption}
\usepackage{subcaption}
\usepackage{multirow}
\usepackage{float}
\usepackage{dcolumn}
\usepackage{bm}
\begin{document}
\title{Gravitational lensing by the hairy Schwarzschild black hole}
\author{Sohan Kumar Jha}
\email{sohan00slg@gmail.com}
\affiliation{Chandernagore College, Chandernagore, Hooghly, West
Bengal, India}
\author{Anisur Rahaman}
\email{anisur@associates.iucaa.in; manisurn@gmail.com
(Corresponding Author)} \affiliation{Durgapur Government College,
Durgapur, Burdwan - 713214, West Bengal, India}

\date{\today}
\begin{abstract}
\begin{center}
Abstract
\end{center}
In this manuscript, we consider the hairy Schwarzschild black hole
that evades the no-hair theorem. The hair is induced by an
additional source from surroundings, such as dark matter, that has
a constant energy-momentum tensor(EMT). We study the strong
gravitational lensing of light in the background of the hairy
Schwarzschild black hole. We observe that the lensing coefficient
$\overline{a}$ increases with $\alpha$ but decreases with
$\ell_0$. The opposite effect is observed for the lensing
coefficient $\overline{b}$ and the impact parameter $b_m$. We also
notice that the angular position $\theta_\infty$ decreases with
$\alpha$ but increases with $\ell_0$, whereas the angular
separation $s$ increases with $\alpha$ and decreases with
$\ell_0$.  For all parameters mentioned, we regain their values
for the Schwarzschild black hole whenever we put either $\alpha=0$
or $\ell_0=1$. With the help of the Gauss-Bonnet theorem, we
briefly describe the weak gravitational lensing in the background
of the hairy Schwarzschild black hole.
\end{abstract}
\maketitle
\section{Introduction}
Deflection of a light ray in the  gravitational field due to
astronomical objects is in general nominated as gravitational
lensing, and the object gives rise to deflection of light is
referred to as  gravitational lens. Gravitational lensing by black
holes is one of the most important and powerful  astrophysical
tools for probing the strong field characteristics of gravity. It
is anticipated to render veritable assessment of implementation of
modified theories of gravity in the strong field regimes and also
in the cosmic censorship hypothesis. Although the  of general
theory of relativity got its impotent operation through the
Gravitational lensing quite a long ago \cite{EINST}, the
advertisement of capturing of image concerning the supermassive
black hole $ M87^*$ at the center of the neighboring elliptical
galaxy $M87$  by the Event Horizon Telescope (EHT) collaboration
is a great triumph in the field of general relativity \cite{ EHT1,
EHT2, EHT3, EHT4, EHT5, EHT6}. In general, the image of black
holes with a girding accretion fragment appears distorted due to
the strong gravitational lensing effect. In this way, black holes
are anticipated to cast murk on the bright background which is
related to the exitances of an event horizon and thus an unstable
photon region \cite{BOZZ}. The shadow and the lensing effect
associated with that is of great significance and scientific
importance. It can help us to probe the geometrical structure of
the event horizon and maybe the externally observable classical
parameters, e.g., mass, electric charge, and angular momentum
through which the black hole is characterized. The full
proposition of gravitational lensing was developed following weak
field approximation scheme and it was plant successful to explain
the physical compliances. Still nearly two and half decades back
the scientific community started looking at the marvels from the
perspective of strong gravitational field \cite{VERG, BAR, FLAKE,
VRIB, FRIT, BOZZ, EIRO, ELL}.

 A semi-analytical treatment was adopted towards the disquisition
about geodesics in Kerr spacetime geometry in the article
\cite{VERG}. In the composition \cite{BAR} the appearance of a
black hole in front of a uniform background was studied. In the
article \cite{FLAKE}, the authors considered the emission of the
accretion inflow as source. In an important composition Virbhadra
and Ellis \cite{VRIB} showed that a source behind a Schwarzschild
black hole would produce one set of infinite relativistic images
on each side of the black hole. These images are produced when
light ray with small impact parameter wind one or several times
around the black hole before it emerges out. Later on, in the
composition \cite{FRIT} gain an exact lens equation through an
indispensable expression of the problem, which shred an integral
expressions as a results, and they compared their results to
earlier work presented in \cite{VRIB}. The same problem has been
delved by Bozza et al. in the article \cite{BOZZ} where a strong
field limit was first defined for the Schwarzschild black hole
lensing effect and it was used analytically to determine the
position and characteristics of all the relativistic images. The
article \cite{EIRO} was devoted with same fashion for the
Reissner-Nordstrom black hole. Lately, in an another composition
\cite{ELL}, Virbhadra and Ellis made an attempt to distinguish the
main features of gravitational lensing by normal black holes and
by naked singularities utilizing the Janis, Newman, Winicour
metric. They asserted that this study may facilitate the probable
test for the  cosmic censorship hypothesis. The gravitational
lansing is not only an important tool in astrophysics to test the
actuality of black holes and to separate between black holes and
other compact objects but also to characterize the matter
distribution at galactic or extragalactic scales. An intriguing
donation of Gibbons and Werner is to establish a link between the
deviation angle with the topology of the  spacetime by means of
the Gauss-Bonnet theorem \cite{ GIBB} where they showed that in
asymptotically flat and static spacetimes the deviation angle can
be attained by integrating the Gaussian optic curve over an
infinite domain of integration immediately outside of the event
horizon.

The gravitational lansing is not only an important tool in
astrophysics to test the existence of  black holes and to
differentiate between  black holes and other compact objects but
also to characterize the matter distribution at galactic or
extragalactic scales. A contribution of Gibbons and Werner is to
establish a  link between the deflection angle with the topology
of the spacetime by means of the Gauss-Bonnet theorem \cite{GIBB}
where they showed that in asymptotically flat and static
spacetimes the deflection angle can be obtained by integrating the
Gaussian optical curvature. High resolution imaging of black holes
by VLBI \cite{ISHIHARA1} could be suitable to detect the
relativistic images and recoup information about strong fields
stored within these new observables.

Alternative theories of gravitation  amenable to strong gravity
must agree with general relativity in the weak field limit, in
order to show deviations from general relativity it is necessary
to probe strong fields in some way. Indeed, deviation of light
rays in strong fields is one of the most favorable grounds where a
theory of gravitation can be tested promisingly. The study of null
geodesics in strong fields is much involved and is generally
computed using numerical techniques. Analytical treatment would
would enlighten the dependence of the observable on the parameters
of the system, allow easy checks about the delectability of the
images and open the way of comparisons between the results in
different metrics. In the article \cite{AGV}, a new way to expand
the deflection angle in the Schwarzschild metric was suggested.
The deflection angle near its divergence was approximated by its
leading order and its first regular term and then plugged into the
the massive body. This result was generalized to stationary
spacetimes by Werner applying the Finsler-Randers geometry
\cite{CMW}. On the other hand, another method was applied by
Ishihara et. al.\cite{ISHIHARA1,ISHIHARA2} in order to calculate
the deflection angle using the Gauss-Bonnet theorem. This method
was generalized to stationary spacetimes by Ono et.al. \cite{ONO1,
ONO2, ONO3}.

Gravitational lensing through black holes commenced to be
observationally important within side the 1990s, which stimulated
numerous quantitative studies of the Kerr metric. Vazquez and
Esteban explored the phenomenology of sturdy subject gravitational
lensing through the usage of a Kerr black hole. They have advanced
a widespread technique to calculate the positions and
magnification of all images for an observer and supply some
distance far far away  from the black hole and at arbitrary
inclinations. Since then, gravitational deflection of mild through
rotating black holes has obtained vast interest because of the
great development of cutting-edge observational facilities. The
black holes would possibly have the exciting function in
assessment to the black hollow without any hair. This might also
additionally assist us to apprehend the hairy black in a higher
dimensin. Recent time witnessed an exiting interest in studyig
gravitational lensing through black holes because of the Event
Horizon Telescope (EHT) observations. This paper pursuits to look
into the sturdy gravitational lensing of hairy Schwarzschild holes
 he purpose of this article is to examine the function of and examine
the phenomenological differences he supermassive black holes Sgr
$A^*$ and $M87^*$. the deformation parameter and the number one
hair on gravitational lensing observables and time postpone among
the relativistic images. Further, thinking about the suprmassive
black holes Sgr $A^*$ and $M87^*$ as cosmological  gravitational
lense, we obtain the positions, separation, magnification, and the
time detention of relativistic images.

The article is organized as follows. Sec. I of this article
contains a discussion on Hairy black hole. In Sec. II we describe
the strong gravitational lansing corresponding to the hairy black
hole considered here. Sec. III is devoted with tee observational
facts in the string gravitational limit conserving Sgr $A^*$ and
$M87^*$ black holes. In Sec V. with  a brief discussion of time
delay we furnish the expected time delay for different black holes
in a tabular form. Weak gravitational lansing scenarios is
considered in Sec. VI. , and the final Sec. VII,  contains the
summary and conclusion of the work

\section{HAIRY SCHWARZSCHILD BLACK HOLES}
In the article \cite{OVALLE1, OVALLE2, OVALLE3}, Ovalle et al.
introduced gravitational decoupling to describe deformations of a
known spherically symmetric solutions induced by additional
sources. The Einstein field equations, in this case, became
\begin{equation}
\tilde{G}_{\mu\nu}=\kappa \left(T_{\mu\nu}+S_{\mu\nu}\right)
\end{equation}
where $T_{\mu\nu}$ and $S_{\mu\nu}$ were the energy momentum
tensors of known solution in general relativity and  the
additional source respectively. In the composition \cite{OVALLE4},
Ovalle et al. considered Schwarzschild black hole surrounded by
spherically symmetric matter. The ansantz for the hairy
Schwarzschild black as used there was
\begin{eqnarray}
ds^{2}&=&-\left(1-\frac{2M}{r}+\alpha e^{-r/(M-\frac{\ell_0}{2})}
\right)dt^{2}+\left(1-\frac{2M}{r}+\alpha e^{-r/(M-\frac{\ell_0}{2})}
\right)^{-1}dr^{2}+r^{2}\left(d\theta^{2}+\sin^{2} \theta d\phi^{2}\right)\label{Metric}\\
&=&-f(r)dt^{2}+f(r)^{-1}dr^{2}+r^{2}\left(d\theta^{2}+\sin^{2}
\theta d\phi^{2}\right).\label{metric}
\end{eqnarray}
were $f(r)=\left(1-\frac{2M}{r}+\alpha
e^{-r/(M-\frac{\ell_0}{2})}\right)$, $M$ represents  the mass of
the black hole and $\alpha$ refers to a parameter that induces
deformation due to additional sources. Here $\ell_{0}=\alpha \ell$
represents the increase in entropy caused by the hair and to
ensure asymptotic flatness, it  must satisfy the condition
$\ell_{0}\leq 2M=\ell_{K}$. In the limit $\alpha \rightarrow 0$
the \ref{Metric} reduces to the standard metric of the
Schwarzschild black hole.

To study gravitational lensing in the background of hairy
Schwarzschild black hole we introduce two dimensionless variables
\begin{equation}
x=\frac{r}{2M}\quad \text{and} \quad
\tilde{\ell_{0}}=\frac{\ell_{0}}{2M}.
\end{equation}
Following articles \cite{BOZZA, BOZZA1, BOZZA2}, we define
$t\rightarrow \frac{t}{2M}$ and  rewrite the metric (\ref{Metric})
in the following form
\begin{equation}
d\tilde{s}^{2}=(2M)^{-2}ds^{2}=-A(x)dt^{2}+A(x)^{-1}dx^{2}
+C(x)\left(d\theta^{2}+\sin^{2} \theta
d\phi^{2}\right),\label{Metric1}
\end{equation}
where
\begin{equation}
A(x)=1-\frac{1}{x}+\alpha e^{-2x/(1-\tilde{\ell_{0}})} \quad \text{and} \quad C(x)=x^{2}
\end{equation}
Horizon radius $x_{h}$ is obtained by solving the equation
\begin{equation}
A(x)=0.
\end{equation}
Let us now plot $A(x)$ versus $x$ for different values $\ell_0$
taking a fixed value  $\alpha=2$, and $x_h$ versus $\alpha$ for
different values $\ell_0$.

\begin{figure}[H]
    \begin{center}
                 \begin{tabular}{cc}
            \includegraphics[scale=0.7]{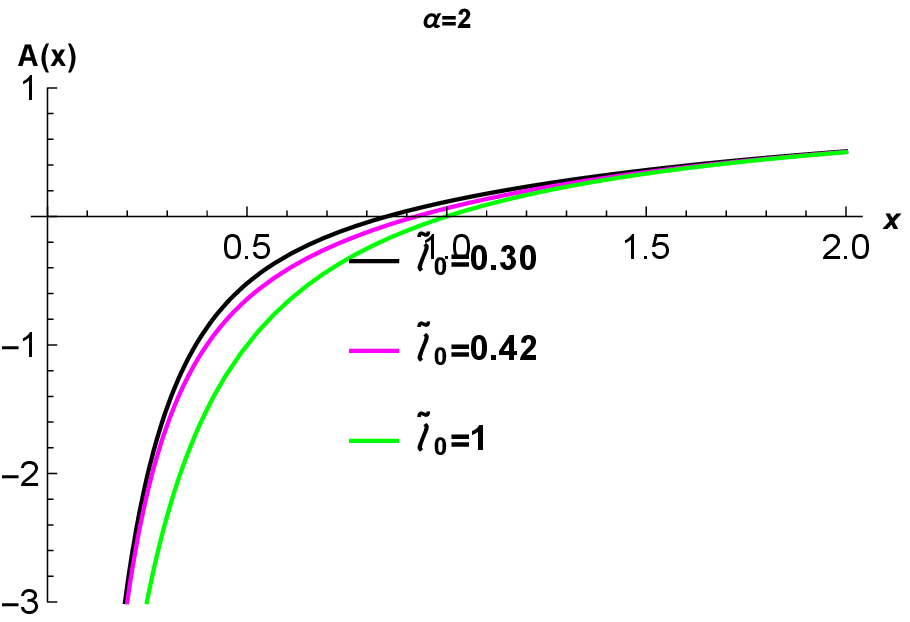}&
                                \includegraphics[scale=0.7]{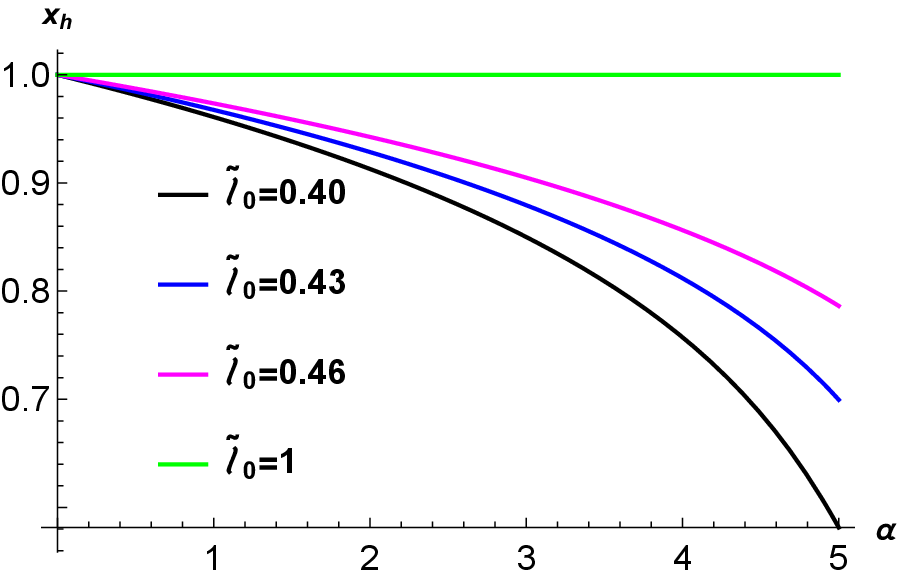}
                 \end{tabular}
        \caption{The left one shows the variation of $A(x)$ with $x$ for
        various values of $\ell_0$ with $\alpha=2$. The right one shows
        the variation of $x_h$ with respect to $\alpha$ for various values of $\ell_0$. }\label{axh}
    \end{center}
\end{figure}
 Fig. \ref{axh} shows that the horizon radius decreases with
$\alpha$, but it increases with $\ell_0$. Note that for $\alpha=0$
that corresponds to $\tilde{\ell}_{0}=1$ we have $x_{h}=1$ which
resembles the Schwarzschild metric.  However when
$\tilde{\ell}_{0}$ is kept constant the horizontal radius
decreases with an increase in $\alpha$, but the radius
 increases with an increase in $\tilde{\ell}_{0}$ when $\alpha$
remains constant. It reflects that the horizontal radius for the
hairy Schwarzschild black hole is smaller than that for the
Schwarzschild black hole. Therefore the presence of hair has a
reducing effect on the Schwarzschild radius. Therefore it is
expected to shows its effect on the entropy, energy emission and
on the shadow. Our objective in this article is to study the
strong gravitational lensing since it has hair strong
gravitational lensing shows a prominent role.
\section{Strong gravitational lensing}
This section is devoted to  study the gravitational deflection of
light in the static spherically symmetric hairy Schwarzschild
 black hole (\ref{Metric1}). Due to spherical symmetry, we can consider
 the propagation of light on the equatorial plane without any
 loss of generality because the same results can be applied to all $\theta$.
 Then the ansantz (\ref{Metric1}) then reduces to
\begin{equation}
d\tilde{s}^{2}=-A(x) dt^{2}+A(x)^{-1} dx^{2}+C(x) d \phi^{2} .
\end{equation}
Since the spacetime is static and spherically symmetric, the
energy $\mathcal{E}=p_{\mu} \xi_{(t)}^{\mu}$ and the angular
momentum $\mathcal{L}=p_{\mu} \xi_{(\phi)}^{\mu}$ remain constant
along the geodesics, where $\xi_{(t)}^{\mu}$ and
${\xi_{(\phi)}^{\mu}}$ are, respectively, the Killing vectors due
to time-translational and rotational invariance \cite{CHANDRA}. So
we have
\begin{equation}
\frac{d t}{d \lambda}=\frac{\mathcal{E}}{A(x)}, \quad \frac{d \phi}{d \lambda}=
-\frac{\mathcal{L}}{C(x)},\label{Conserved}
\end{equation}
where $\lambda$ is the afine parameter along the geodesics. From
Eqn.(\ref{Conserved}) we obtain the following equation for the
null geodesics
\begin{equation}
\left(\frac{d x}{d \lambda}\right)^{2} \equiv \dot{x}^{2}
=\mathcal{E}^{2}-\frac{\mathcal{L}^{2} A(x)}{C(x)}.
\end{equation}
Thus the effective potential $V_{\text {eff }}(x)$ comes out to be
\begin{equation}
V_{\text{eff}}(x)=\frac{\mathcal{L}^{2} A(x)}{C(x)}
=\frac{\mathcal{L}^{2}}{x^{2}}\left(1-\frac{1}{x}+\alpha
e^{-2x/(1-\tilde{\ell_{0}})}\right).
\end{equation}
For circular photon orbits of radius $x_{m}$ the effective
potential has to satisfy the condition
\begin{equation}
\frac{d V_{\text{eff}}}{d x}=0,
\end{equation}
which yields
\begin{equation}
\frac{A^{\prime}(x)}{A(x)}=\frac{C^{\prime}(x)}{C(x)}.
\end{equation}
With the use of the above equation we have
\begin{equation}
\left(2x-3\right)(1-\tilde{\ell_{0}})+2\alpha
e^{(-2x/(1-\tilde{\ell_{0}}))}\left(x(1-\tilde{\ell_{0}})
+x^{2}\right)=0.
\end{equation}
Note that at $x=x_{m}$ the condition  $\frac{d^{2}
V_{\text{eff}}}{dx^{2}}<0$ is maintained. Thus these orbits are
unstable against small radial perturbations. Photons from the far
distance source approach the black hole with
 some impact parameter $b$ with a minimum distance $x_{0}$ and get
  deflected symmetrically to infinity. The impact parameter $b$
  and the minimum distance $x_{0}$ are related to each other through the equation
\begin{equation}
V_{\text {eff }}(x)=\mathcal{E}^{2} \quad \Rightarrow \quad b \equiv \frac{\mathcal{L}}{\mathcal{E}}
=\sqrt{\frac{C\left(x_{0}\right)}{A\left(x_{0}\right)}}\label{b}
\end{equation}
The impact parameter $b_{m}$ corresponding to  $x_{0}=x_{m}$.

\begin{figure}[H]
    \begin{center}
                 \begin{tabular}{cc}
            \includegraphics[scale=0.7]{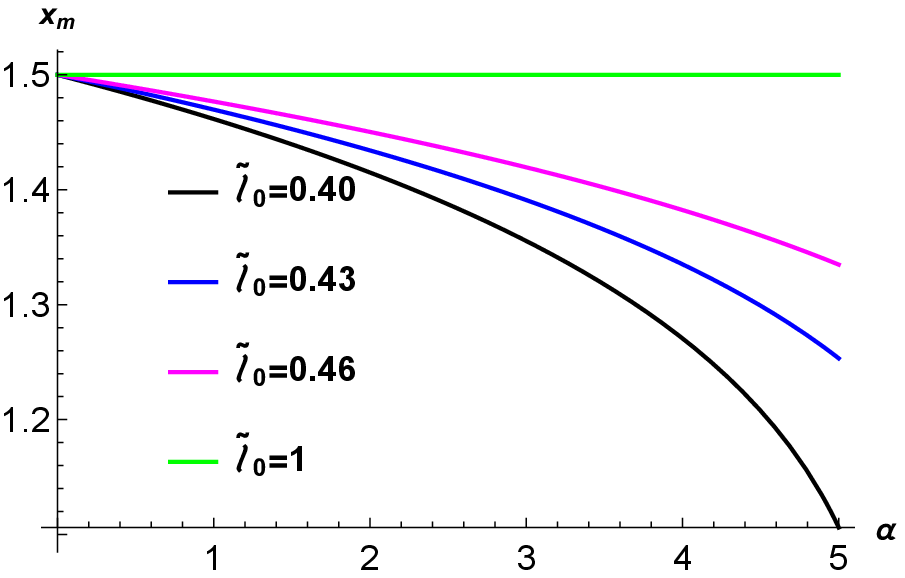}&
                                \includegraphics[scale=0.7]{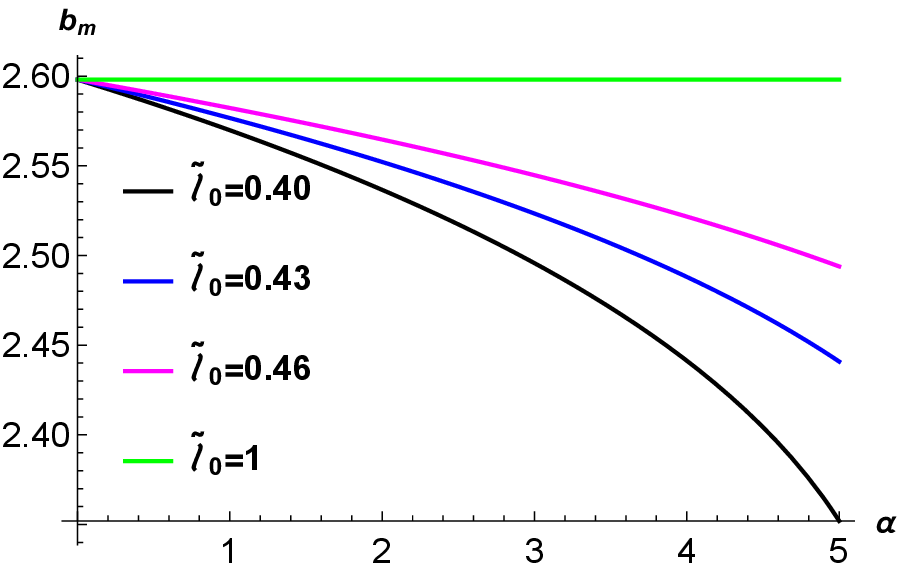}
                 \end{tabular}
        \caption{Plot showing variation of $x_m$ and $b_m$ with respect to $\alpha$ for various values of $\ell_0$. }\label{xm}
    \end{center}
\end{figure}
We can infer from the Fig.\ref{xm},  that the radius $x_m$ and the
impact parameter $b$ at $x_m$ i.e $b_m$ decrease with $\alpha$ but
increase with $\ell_0$. Let us look towards the gravitationsl
deflection angle of light as has been formulated in the article
\cite{VIRBHADRA, WEINBERG}
\begin{equation}
\alpha_D\left(x_{0}\right)=I\left(x_{0}\right)-\pi,\label{Deflection}
\end{equation}
where
\begin{equation}
I\left(x_{0}\right)=\int_{x_{0}}^{\infty}\frac{2}{\sqrt{A(x)C(x)}\sqrt{\frac{A(x_{0})C(x)}{C(x_{0})A(x)}-1}}dx\label{I0}
\end{equation}
In the absence of black hole total change in $\phi$ is $\pi$ as
photons follow straight line trajectory and hence, by Eqn.
(\ref{Deflection}) the deflection angle becomes zero. The
deflection angle $\alpha\left(x_{0}\right)$ increases as the
impact parameter $b$ increases and at some point, exceeds $2\pi$
resulting in complete loops of photon around the black hole.
 When $x_{0}=b_{m}$ i.e the minimum distance equals the radius
 of photon sphere, the deflection angle diverges. For $b<b_{m}$,
 photons get captured. In the strong field limit light rays
 pass close to the black hole. Following the method developed  in
 \cite{BOZZA}, we expand the deflection angle about the photon
 sphere where the angle diverges. Here, we introduce the variable
  $z=1-x_{0}/x$ in a similar way it was found to use in \cite{GHOSH}.
  Thus, the integral (\ref{I0}) can be re-written as
\begin{eqnarray}
I(x_0) &=& \int_{0}^{1}R(z,x_0) f(z,x_0) dz,\label{def1}
\end{eqnarray}
with
\begin{eqnarray}
R(z,x_0) &=&  \frac{2x^{2}\sqrt{C(x_0)}}{x_{0}C(x)},\\\label{RX}
f(z,x_0) &=& \frac{1}{\sqrt{A(x_0)-\frac{A(x)}{C(x)}C(x_0)}}\label{FX},
\end{eqnarray}
The function $R(z,x_{0})$ is regular for all values of $z$ and
$x_{0}$ whereas the function $f(z,x_0)$ diverges when
$z\rightarrow 0$. To bypass divergence  we perform the Taylor
series expansion of the expression contained within the square
root of the function (\ref {FX}) and approximated appropriately to
get rid of the divergences and we have
\begin{eqnarray}
f_0(z,x_0) &=& \frac{1}{\sqrt{a_{1}(x_0) z + a_{2}(x_0) z^2}},
\end{eqnarray}
where
\begin{eqnarray}
a_{1}(x_0) &=& \frac{x_{0}}{C(x_0)}\left[\left(C^\prime(x_0) A(x_0)-A^\prime(x_0) C(x_0)\right)\right],\\
a_{2}(x_0) &=& \frac{1}{2}\left[\frac{(2x_{0}C(x_{0})-2x_{0}^{2}C^{\prime}(x_{0}))(C^\prime(x_0) A(x_0)-A^\prime(x_0) C(x_0))}{c^{2}(x_{0})}+\frac{x_{0}}{C(x_{0})}\left(C^{\prime\prime}(x_{0})A(x_{0})-A^{\prime\prime}(x_{0})C(x_{0})\right)\right]
\end{eqnarray}
With above definitions, we split the integral (\ref{def1}) into
two following the development in \cite{BOZZA}
\begin{eqnarray}
I(x_0) &=& I_D(x_0)+I_R (x_0),
\end{eqnarray}
where
\begin{eqnarray}
I_D(x_0) &=& \int_0 ^1 R(0,x_m)f_0(z,x_0)dz,\label{divergent}\\
I_R(x_0) &=& \int_0 ^1
\Big(R(z,x_0)f(z,x_0)-R(0,x_m)f_0(z,x_0)\Big)dz.\label{regular}
\end{eqnarray}
When $x_0=x_m$, we have $a_1(x_0)=0$ and hence,
$f_0(z,x_0)=1/\sqrt{a_2(x_m)}z$. This makes the integral
(\ref{divergent}) divergent at $x_0=x_m$. The exact solution of
$I_D(x_0)$ is given by
\begin{equation}
I_D(x_0)=R(0,x_m)\frac{2}{\sqrt{a_{2}(x_0)}}\log\frac{\sqrt{a_{2}(x_0)}
+\sqrt{ a_{1}(x_0)+a_{2}(x_0)}}{ \sqrt{a_{1}(x_0)}}.\label{ID}
\end{equation}
we expand $a_1(x_0)$ in the following fashion
\begin{equation}
a_1(x_0)=\frac{2x_{m}a_2(x_m)}{C(x_m)}\left( x_0-x_m \right) +
O\left( x_0-x_m \right)^2,
\end{equation}
and retained the terms up to $O(x_0-x_m)$. With this, we can
re-write the integral (\ref{ID})  as
\begin{equation}
I_D(x_0)=-a \log
\left(\frac{x_0}{x_m}-1 \right)+b_D+ O(x_0-x_m), \label{ID1}
\end{equation}
where
\begin{eqnarray}
&& a=\frac{R(0,x_m)}{\sqrt{a_2(x_m)}} \label{a} \\ %
&& b_D=\frac{R(0,x_m)}{\sqrt{a_2(x_m)}} \log \frac{2C(x_m)}{x_{m}^{2}} \label{bD}.
\end{eqnarray}
Now, we expand Eqn. (\ref{regular}) in powers of $x_0-x_m$ and
retain terms up to $O(x_0-x_m)$, we will have
\begin{eqnarray}
I_R(x_0)&=&\int\limits_0^1 g(z,x_m) dz +O(x_0-x_m) \\
&=& b_R+O(x_0-x_m), \label{IR}
\end{eqnarray}
where
\begin{equation}
b_R=I_R(x_m) \label{bR}.
\end{equation}
Next, we expand the Eqn. (\ref{b}) about $b_m$ which yields
\begin{equation}
b-b_m=P\left(x_0-x_m\right)^2, \label{expand}
\end{equation}
where
\begin{equation}
P=\frac{a_2(x_m)}{2x_m^2}\sqrt{\frac{C(x_m)}{A(x_m)^3}}.
\end{equation}
With the help of equations (\ref{ID1}) (\ref{IR}) and
(\ref{expand}), the angle of deflection as a function of $b$ are
found out to be
\begin{eqnarray}
&& \alpha_D(b)=-\overline{a} \log \left( \frac{b}{b_m} -1
\right) +\overline{b}+O(b-b_m) ,\label{alphab}\\%
\text{where}\\\nonumber
&& \overline{a}=\frac{a}{2}=\frac{R(0,x_m)}{2\sqrt{a_2(x_m)}},  \label{afinal}\\%
&& \overline{b}=-\pi+b_R+\overline{a}\log{\frac{2C^2(x_m)a_2(x_m)}{A(x_m)x_m^4}}.\label{bfinal}
\end{eqnarray}

\begin{figure}[H]
    \begin{center}
        \begin{tabular}{c c}
            \includegraphics[scale=0.7]{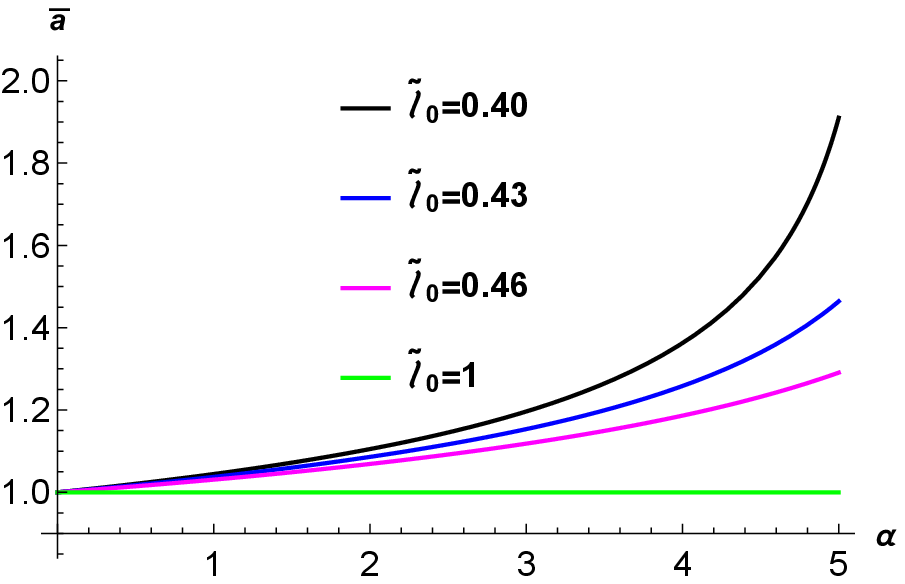}&
            \includegraphics[scale=0.7]{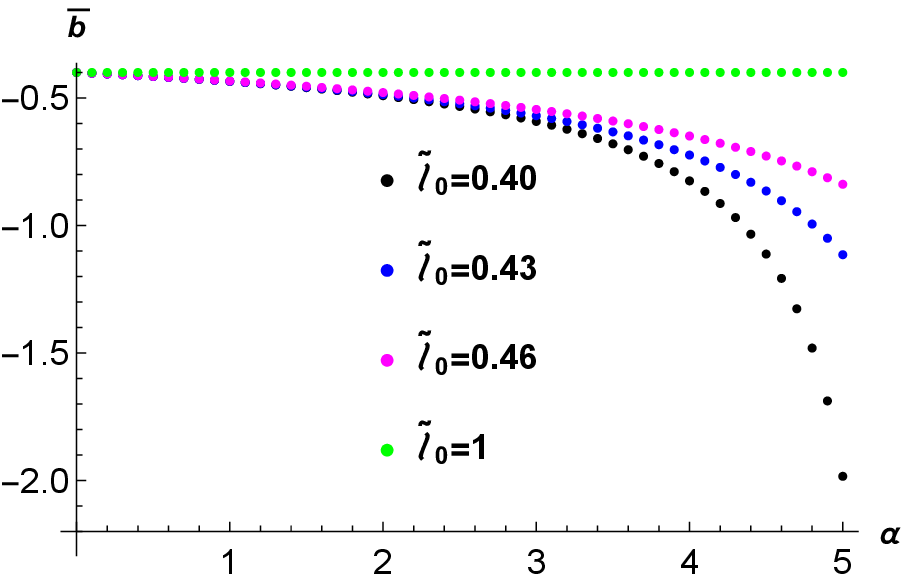}\\
        \end{tabular}
        \caption{Plot showing the variation of strong lensing
        coefficients $\overline{a}$ and $\overline{b}$ with respect to
         $\alpha$ for various values of $\ell_0$. }\label{coefficient}
    \end{center}
\end{figure}

A careful look on the Fig. \ref{coefficient} and the table I
reveals that the coefficient $\overline{a}$ increases with
$\alpha$ but decreases with $\ell_0$. The coefficient
$\overline{b}$, on the other hand, decreases with $\alpha$ and
increases with $\ell_0$. The green line signifies the fact that
when $\ell_0\rightarrow1$, the ansantz (\ref{Metric}) reduces to
that for the Schwarzschild black hole. In the following Fig 4. the
variation of deflection angle in strong gravitation lensing with
respect to $\alpha$ and $b$ are depicted.
\begin{figure}[H]
    \begin{center}
        \begin{tabular}{c c}
            \includegraphics[scale=0.7]{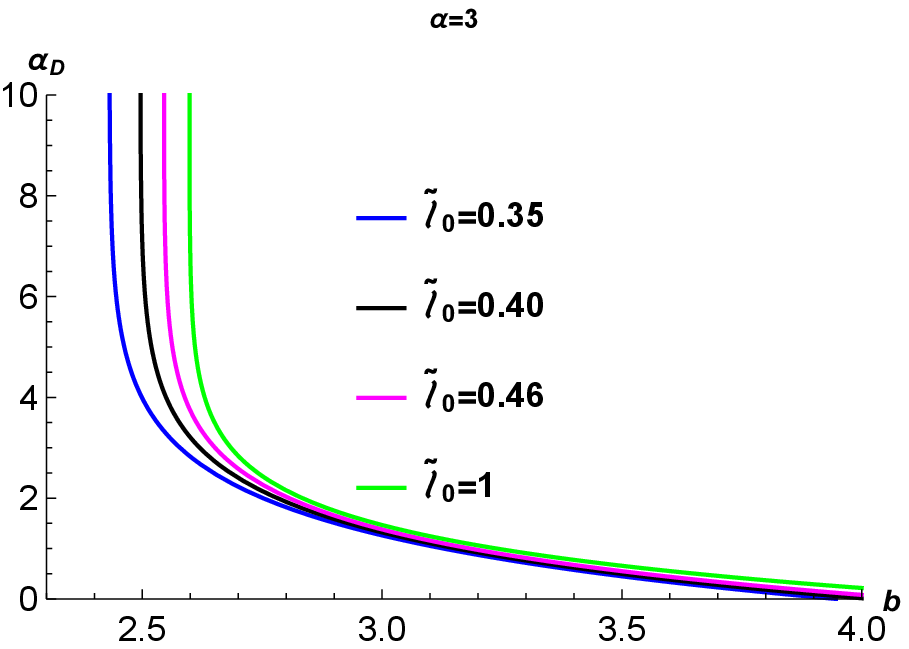}&
            \includegraphics[scale=0.7]{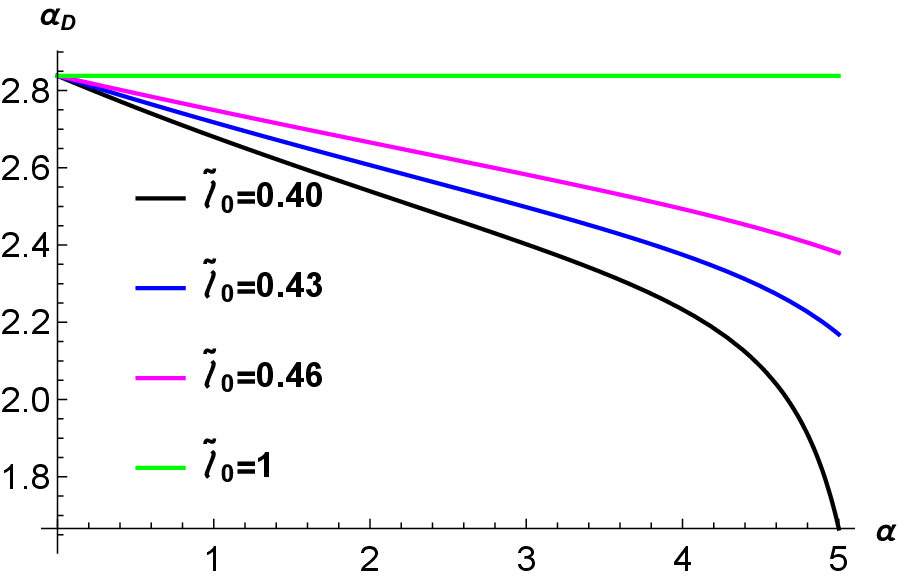}\\
        \end{tabular}
        \caption{The left one shows the variation of the deflection angle $\alpha_D$
        with respect to the impact parameter $b$ for different values of $\ell_0$ with $\alpha=3$. The right one shows the variation of the deflection angle with respect to $\alpha$ with $b=2.7$. }\label{strong}
    \end{center}
\end{figure}
It would be beneficial to compare the lensing coefficients in the
presence  of hair and and the absence of it in connection to the
Schwarzschild black. It is furnished below in a tabular form
\begin{table}[!htp]
\centering
\setlength{\tabcolsep}{0pt}
\begin{tabular*}{\textwidth}{@{\extracolsep{\fill}\quad}lccccccc}
\hline\hline\\
{$\alpha$ } & {$\ell_0$} & {$\bar{a}$} & { {$\bar{b}$}} &$b_m/R_s$ & {$\delta\bar{a}$} & { {$\delta\bar{b}$}} &$\delta u_m/R_s$
\\ \hline
\hline
\\
\multirow{3}{*}{1}& 0.40 &1.04441& -0.435351& 2.56985&0.0444121& -0.035121& -0.0282291 \\
&0.50 &1.02268& -0.429227,&2.58807&0.0226802& -0.028997&-0.0100077 \\
&0.60 &1.0077& -0.414335& 2.5959&0.00769851& -0.0141049& -0.00218009 \\
\hline \\
\multirow{3}{*}{2} &0.40 &1.10507& -0.491267& 2.53661&0.105066& -0.0910372& -0.0614685 \\
&0.50 &1.04924& -0.465694& 2.57728&0.0492353& -0.0654637& -0.0207955 \\
&0.60 &1.01582& -0.429526& 2.59366&0.0158171& -0.0292962& -0.00441136 \\
\hline \\
\multirow{3}{*}{3} & 0.40 &1.19637&-0.592265& 2.49574&0.196367& -0.192035&-0.102341  \\
& 0.50 &1.08101& -0.512881& 2.56555&0.0810136& -0.112651& -0.0325222\\
& 0.60 &1.0244& -0.445935& 2.59138&0.0243984&-0.0457052& -0.00669681 \\
  \hline\hline
\end{tabular*}
\caption{The lensing coefficients for the hairy Schwarzschild
black holes and deviation of these coefficients from Schwarzschild
black holes ($\alpha=0$). Here $\delta$ indicates the difference
between the presence of hair and in its absence.} 
\end{table}

The deflection angle as a function of the angular separation of
the image from lens $\theta$ can be written as
\begin{equation}
\alpha_D(\theta)=-\overline{a} \log \left( \frac{\theta D_{OL}}{b_m} -1\right)
+\overline{b}+O(b-b_m)\label{alphat}
\end{equation}
where $ D_{OL}$ is the distance between the observer and the black hole.
\section{OBERVABLES IN THE STRONG FIELD LIMIT}
In this section, with help of Eqn. (\ref{alphat}) and the lens
equation derived in the article \cite{BOZZA1}, we make an attempt
to  derive the expressions for the position and magnification of
relativistic images. The knowledge gained  from the article
\cite{BOZZA1} enables us to write the lens equation:
\begin{equation}
\beta=\theta-\frac{D_{LS}}{D_{OS}} \Delta \alpha_n.%
\label{Lens}
\end{equation}
where $D_{LS}$ is the lens-source distance, $D_{OS}=D_{OL}+D_{LS}$
 is the observer-source distance,  $\beta$ is the angular separation
 between the source and the lens, and $\Delta \alpha_n =\alpha(\theta)
-2n \pi$ is the offset of the deflection angle, n being the positive
integer number that gives the winding number of loops around the black hole.\\
To calculate the offset angle, we first need to find the
angle $\theta_{n}^{0}$ that solves the equation $\alpha(\theta)=2n\pi$.
The solution is given by
\begin{eqnarray}
&& \theta_n^0=\frac{b_m}{D_{OL}} \left(1+e_n \right) \label{theta0}\\%
&& e_n=e^{\frac{\overline{b}-2n\pi}{\overline{a}}}.\label{en}
\end{eqnarray}
Now, we consider $\Delta\theta_n= \theta-\theta_n^0$ and
expand $\alpha(\theta)$ around $\theta=\theta_n^0$ for the
purpose of calculating the offset angle. It yields
\begin{equation}
\Delta \alpha_n=-\frac{\overline{a} D_{OL}} {b_m e_n}
\Delta \theta_n .%
\label{deltaalphan}
\end{equation}
If we use the above equation and put
$\theta=\theta_{n}^{0}+\Delta\theta_n$ in Eqn. (\ref{Lens}) we get
\begin{equation}
\beta=\theta_n^0+\Delta \theta_n + \left( \frac{
\overline{a} D_{OL}}{b_m e_n }\frac{D_{LS}}{D_{OS}} \right) \Delta
\theta_n.%
\label{lens new}
\end{equation}
We can neglect the second term in above equation in comparison
to the last term, since $b_m<<D_{OL}$. Thus, the position of
the $n^{\text{th}}$ image is given by \cite{BOZZA}
\begin{equation}
\theta_n = \theta_n^0 +\frac{ b_m e_n\left(\beta-\theta_n^0\right)
D_{OS}}{\overline{a} D_{LS}D_{OL}}.
\label{Images}
\end{equation}
When the image position and the source position coincides, we get
$\beta=\theta_{n}^{0}$ and hence, using the above equation we get
$\theta_n = \theta_n^0$. It implies that the correction to the
$n^{\text{th}}$ image position becomes zero. the Eqn.
(\ref{Images}) gives the image position when the source and the
image are on the same side. To get image position on the other
side of the source, we need to replace $\beta$ by $-\beta$.

In addition to the source position, magnification of the images is
another good source of information \cite{BOZZA}. It is given by
\begin{eqnarray}
\mu_n &=& \left(\frac{\beta}{\theta} \;
\;\frac{d\beta}{d\theta} \Bigg|_{\theta_n ^0}\right)^{-1}\\
&=&e_n \frac{ b_m^2\left(1+e_n \right)
D_{OS}}{\overline{a} \beta D_{OL}^2 D_{LS}}\label{magnification}
\end{eqnarray}
It is evident from the above equation that the magnification
decreases with increase in $n$. Next, we consider the
 simplest scenario where the outermost single-loop image
  $\theta_1$ is resolved as a single image, while all the
  other inner packed images are clubbed together
  at $\theta_{\infty}$. Three observables are given by
\begin{eqnarray}
\theta_\infty &=& \frac{b_m}{D_{OL}},\\
s &=& \theta _1-\theta _\infty = \theta_\infty \;\; e^{\frac{\bar{b}-2\pi}{\bar{a}}},\label{s}\\
r &=&  \frac{\mu_1}{\sum{_{n=2}^\infty}\mu_n } = e^{\frac{2 \pi}{\bar{a}}},\qquad r_{\text{mag}}= 2.5 \log(r) = \frac{5\pi}{\bar{a}\ln 10}.\label{obs}
\end{eqnarray}
Here, $\theta_{\infty}$ gives the position of the innermost packed
images, $s$ gives the angular separation between the first image
and the others, and $r$ gives the ratio of the flux from the first
image to those from the remaining images. It must be noted that,
whereas $\theta_{\infty}$ and $s$ depend on the black hole's
distance from the observer, $r_{\text{mag}}$ does not depend on
the distance.

 Now, we consider supermassive black holes $Sgr A*$
and $M87*$. We calculate lensing observables for these black holes
using the ansantz (\ref{Metric}). The mass and distance from the
Earth of  $Sgr A*$ and $M87*$ are, respectively, given by
$M=3.98\times 10^6 M_{\odot}$ and $D_{OL}=7.97$ kpc for Sgr A*
\cite{TD} and $M=(6.5\pm 0.7)\times 10^9 M_{\odot}$ and
$D_{OL}=(16.8\pm 0.8)$ Mpc for M87* \cite{EHT1, EHT2}.

\begin{figure}
    \begin{center}
        \begin{tabular}{c c}
            \includegraphics[scale=0.84]{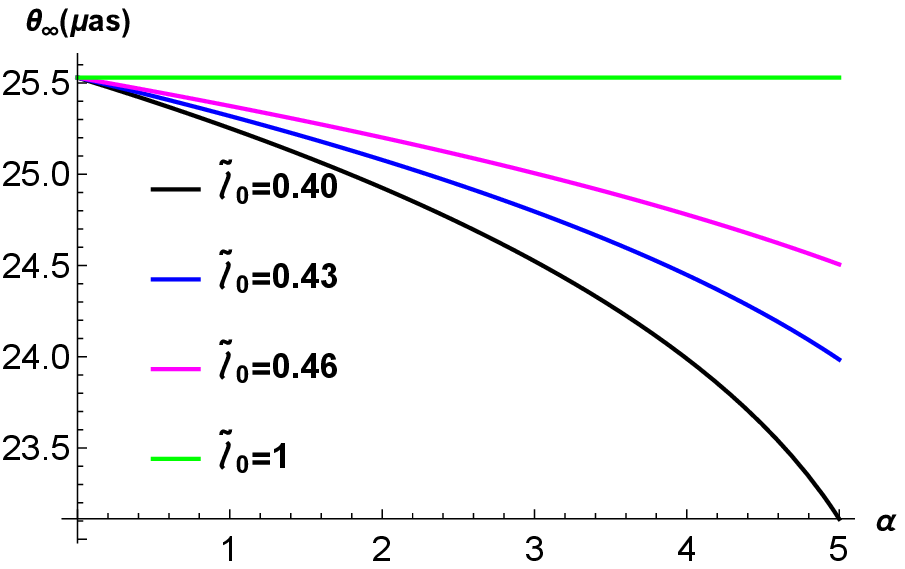}&
            \includegraphics[scale=0.84]{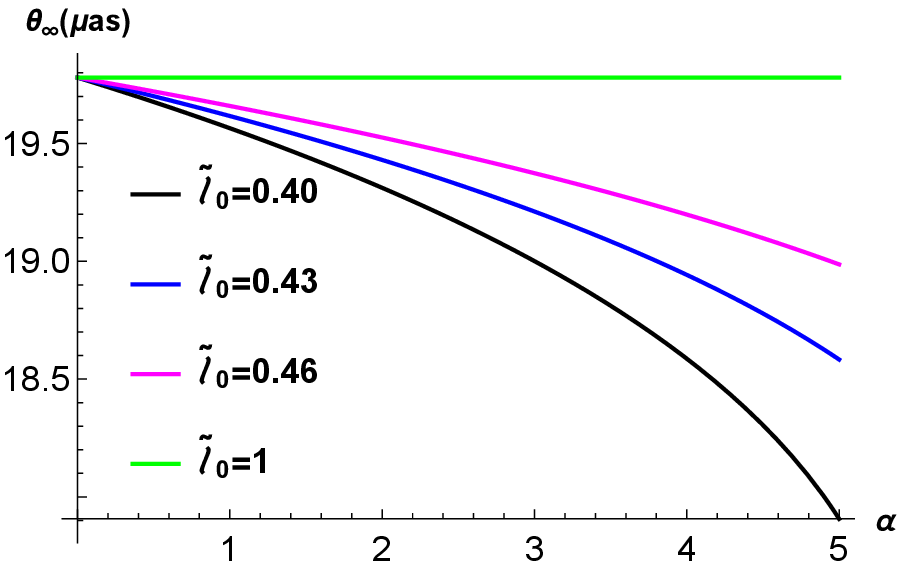}\\
            \includegraphics[scale=0.84]{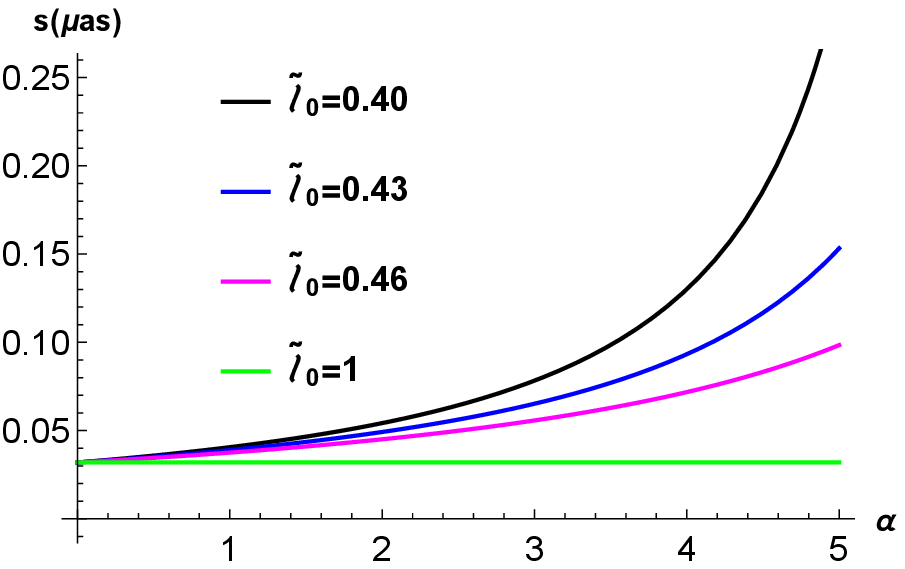}&
                       \includegraphics[scale=0.84]{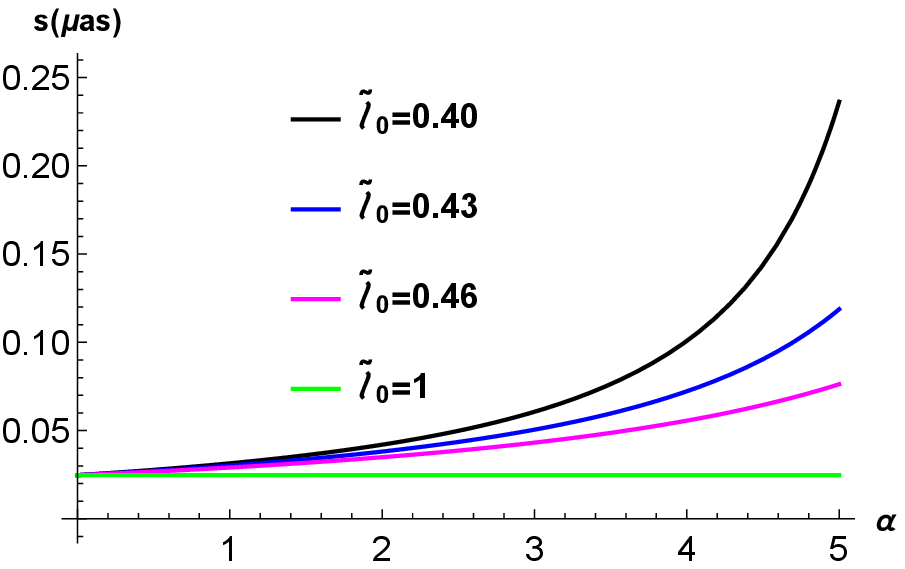}\\
    \multicolumn{2}{c}{\includegraphics[scale=0.84]{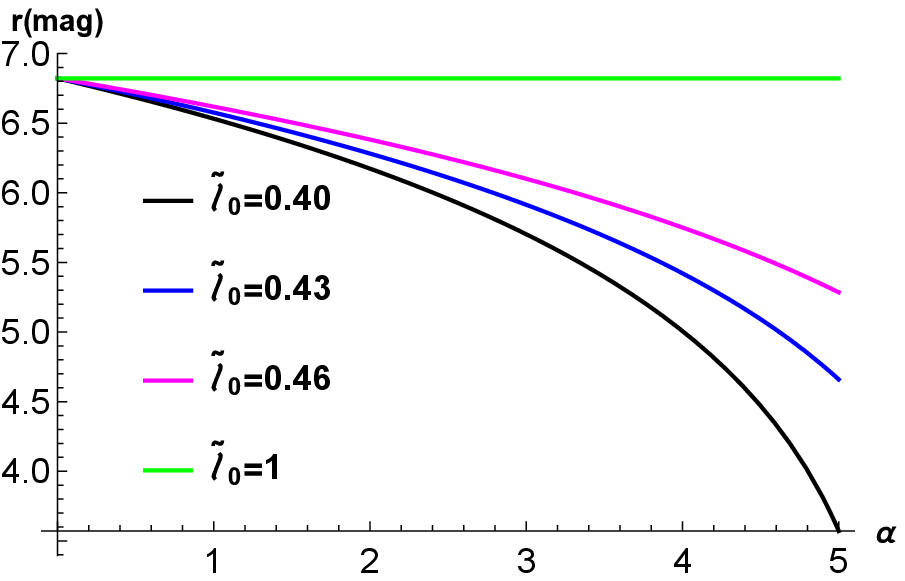}}
        \end{tabular}
        \caption{Lensing observables, $\theta_\infty$, $r(mag)$,
         and $s$ as a function of $\alpha$ for Sgr A* (left panel)
          and M87* (right panel) black holes. }\label{observe}
    \end{center}
\end{figure}

The Fig.\ref{observe} demonstrate the variation of lensing
observables with respect to $\alpha$ for different values of
$\ell_0$ for the Sgr A* and M87* black holes. We can
 infer from the plots that $\theta_\infty$ and $r(mag)$
 decrease with an increase in $\alpha$, whereas these two
 increase with an increase in $\ell_0$. The lensing
 observable $s$ increases with respect to $\alpha$
 and decreases with respect to $\ell_0$. The table
 below shows some values of lensing obserevables for
 the Sgr A* and M87* black holes.

\begin{table}[H]
 \begin{centering}
\setlength{\tabcolsep}{0pt}
    \begin{tabular*}{\textwidth}{@{\extracolsep{\fill}\quad}lccccccc }
\hline\hline
\multicolumn{2}{c}{}&
\multicolumn{2}{c}{Sgr A*}&
\multicolumn{2}{c}{M87*}& \\
{$\alpha$ }& {$\ell_0$}  & {$\theta_\infty $ ($\mu$as)} & {$s$ ($\mu$as) } & {$\theta_\infty $ ($\mu$as)} & {$s$ ($\mu$as) } & {$r_{mag} $ }
\\ \hline
\hline
\\

\multirow{3}{*}{1}&0.40&25.2524& 0.0406027& 19.5651& 0.0314583& 6.53179 \\
&0.50&25.4314& 0.0358804& 19.7038& 0.0277995& 6.67059\\
&0.60&25.5084& 0.033129& 19.7634& 0.0256677& 6.76976\\
\hline \\
\multirow{3}{*}{2}&0.40&24.9258& 0.0542333& 19.312& 0.042019& 6.17328 \\
&0.50&25.3254& 0.0407468& 19.6217& 0.0315698& 6.50177\\
&0.60&25.4864& 0.0343882& 19.7464& 0.0266433& 6.71566\\
\hline \\
\multirow{3}{*}{3}&.040 &24.5241& 0.078294& 19.0008& 0.0606608& 5.70217 \\
&0.50&25.2102& 0.0469109& 19.5324& 0.0363457& 6.31063\\
&0.60&25.464& 0.0357364& 19.729& 0.0276879& 6.6594\\
\hline\hline
\end{tabular*}
\end{centering}
\caption{Strong-lensing observables for the black hole $Sgr A*$
and $M87^*$. \label{table2}}
\end{table}

We obtain the Einstein ring when we have perfect alignment i.e $\beta=0$ \cite{EINSTEIN}. Solving the
 Eqn. (\ref{Images}) for $\beta=0$ we get
\begin{equation}
\theta_n^{E} = \left(1-\frac{ b_m e_nD_{OS}}{\overline{a} D_{LS}D_{OL}}\right)\theta_n^0 .\label{Einstein}
\end{equation}
Considering the case where the lens is midway between source and
obserber i.e $D_{OS}=2D_{OL}$ and taking $D_{OL}>>b_{m}$ we obtain
from Eqn. (\ref{Einstein})
 \begin{equation}
\theta_n^{E} = \frac{b_m}{D_{OL}}\left(1=e_n\right) .\label{Einstein1}
\end{equation}
Which gives the angular radius of the $nth$ relativistic Eistein ring. For $n=1$ we display Einstein rings for Sgr A* and M87*.
\begin{figure}[H]
    \begin{center}
        \begin{tabular}{c@{\hspace{2em}}c}
            \includegraphics[scale=0.7]{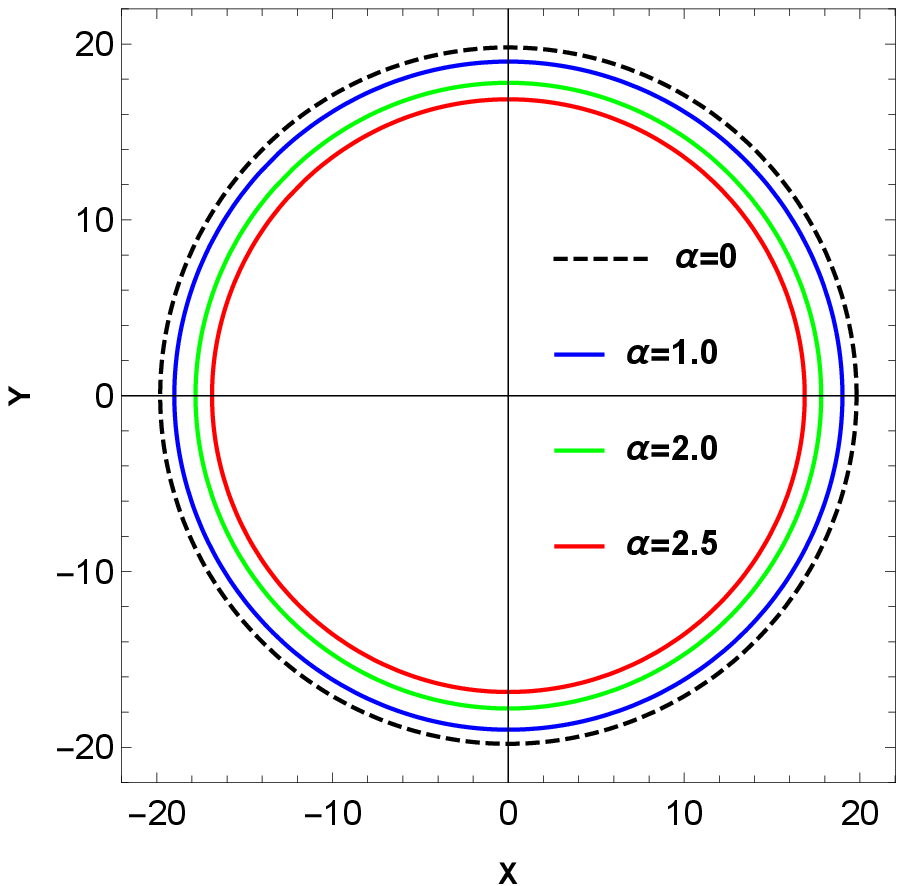}&
            \includegraphics[scale=0.7]{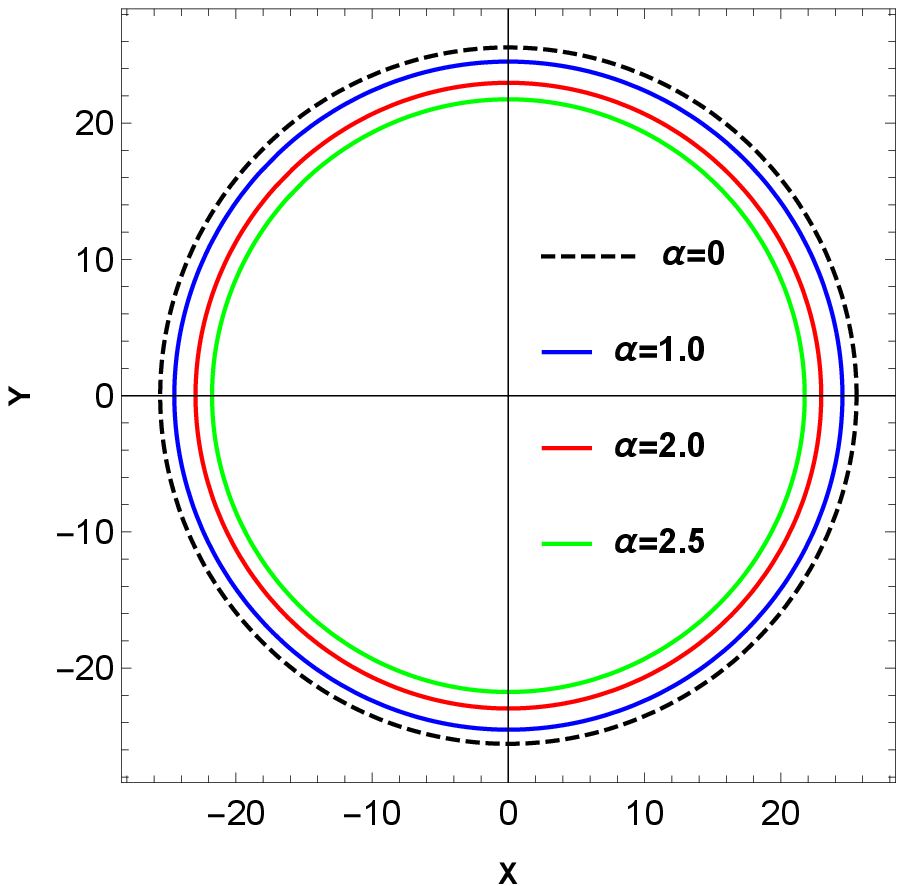}\\
        \end{tabular}
        \caption{Plots showing the outermost Einstein rings. The left one is for $M87^{*}$ and the right one is for $Sgr A^{*}$. Here we have taken $\ell_{0}=0.2$  }\label{Rings}
    \end{center}
\end{figure}
Fig. (\ref{Rings}) shows that the radius decreases with an
increase in $\alpha$ for both the black holes.

\section{Time delay}
In this section we study the time delay between relativistic
images for various black holes. Here we have followed the method
given by Bozza and Manchini \cite{BOZZA2}.  The time delay causes
due to the different paths that photons take while winding the
black hole. The article \cite{BOZZA2} suggests that the time taken
by a photon to reach an observer at infinity from the source is
given by
\begin{eqnarray}\label{TD1}
\tilde{T}(u) &=& \tilde{a} \log\left(\frac{b}{b_m} -1\right) + \tilde{b} +\mathcal{O}(b-b_m),
\end{eqnarray}
Using the above equation we can calculate the time difference between  two  relativistic images. The images are highly demagnified and the separation between the images is of the order of $\mu$as. Here we consider the time delay between the first and second relativistic image  assuming that the two images are on the same side of the source. The time delay is given by \cite{BOZZA2}
\begin{eqnarray}\label{deltaT}
\Delta T_{2,1} &=& 2\pi b_m = 2\pi D_{OL} \theta_{\infty}.
\end{eqnarray}
It is useful at this stage to study the time delay corresponding
to different black holes sitting at the center of the neighboring
galaxies. in this context we furnish the time delay of few black
holes in tabular form in the following table III.
\begingroup
\begin{table*}
    \caption{ Estimation of  time delay for various black holes at the center of nearby  galaxies. Mass ($M$) and distance ($D_{OL}$) are given in the units of solar mass and Mpc, respectively. Time Delays are expressed in minutes. Here we have taken $\ell_{0}=0.2$.
    }\label{Timedelay}
    \begin{ruledtabular}
        \begin{tabular}{c c c c c c c}
Galaxy & $M( M_{\odot})$ & $D_{OL}$ (Mpc) & $M/D_{OL}$ & $\Delta T_{2,1}$& $\Delta T_{2,1}$ & $\Delta T_{2,1}$\\
 &  &  &  & $\alpha=0$ & $\alpha=1.0$ & $\alpha=1.5$ \\
                \hline
Milky Way& $  4.3\times 10^6     $ & $0.0083 $ &       $2.471\times 10^{-11}$ & $11.4967 $ & 11.0193 & 10.702     \\
M87&$ 6.5\times 10^{9} $&$ 16.8 $
&$1.846\times 10^{-11}$& $17378.7 $ & 16657.1  & 16177.5\\
NGC 4472 &$ 2.54\times 10^{9} $&$ 16.72 $
&$7.246\times 10^{-12}$& $6791.06$ & 6509.07 & 6321.66 \\

NGC 1332 &$ 1.47\times 10^{9} $&$22.66  $
&$3.094\times 10^{-12}$& $3930.26$ & 3767.06 & 3658.6 \\

NGC 4374 &$ 9.25\times 10^{8} $&$ 18.51 $
&$2.383\times 10^{-12}$& $2473.12$ & 2370.43 & 2302.18 \\

NGC 1399&$ 8.81\times 10^{8} $&$ 20.85 $
&$2.015\times 10^{-12}$& $2355.48$ & 2257.67 & 2192.67\\

NGC 3379 &$ 4.16\times 10^{8} $&$10.70$
&$1.854\times 10^{-12}$& $1112.24$ &  1066.05 & 1035.36\\

NGC 4486B &$ 6\times 10^{8} $&$ 16.26 $
&$1.760\times 10^{-12}$ & $1604.19$ &   1537.58 & 1493.3\\

NGC 1374 &$ 5.90\times 10^{8} $&$ 19.57 $ &$1.438\times 10^{-12}$& $1577.45$ &  1511.95 & 1468.42 \\

NGC 4649&$ 4.72\times 10^{9} $&$ 16.46 $
&$1.367\times 10^{-12}$& $12619$ & 12095.6 & 11747.3 \\

NGC 3608 &$  4.65\times 10^{8}  $&$ 22.75  $ &$9.750\times 10^{-13}$& $1243.25$ &  1191.62 & 1157.31 \\

NGC 3377 &$ 1.78\times 10^{8} $&$ 10.99$
&$7.726\times 10^{-13}$ & $475.909$ & 456.148 & 443.014 \\

NGC 4697 &$  2.02\times 10^{8}  $&$ 12.54  $ &$7.684\times 10^{-13}$& $540.077$ &   517.651 & 502.746 \\

NGC 5128 &$  5.69\times 10^{7}  $& $3.62   $ &$7.498\times 10^{-13}$& $152.131$ &  145.813 & 141.615 \\

NGC 1316&$  1.69\times 10^{8}  $&$20.95   $ &$3.848\times 10^{-13}$& $451.816 $ &  433.084 & 420.614 \\

NGC 3607 &$ 1.37\times 10^{8} $&$ 22.65  $ &$2.885\times 10^{-13}$& $366.265 $ &  351.08 & 340.971 \\

NGC 4473 &$  0.90\times 10^{8}  $&$ 15.25  $ &$2.815\times 10^{-13}$& $240.628$ &  230.636 & 223.996 \\

NGC 4459 &$ 6.96\times 10^{7} $&$ 16.01  $ &$2.073\times 10^{-13}$ & $186.086 $ &   178.359 & 173.223 \\

M32 &$ 2.45\times 10^6$ &$ 0.8057 $
&$1.450\times 10^{-13}$ & $6.5504 $ &  6.27844 & 6.09766\\

NGC 4486A &$ 1.44\times 10^{7} $&$ 18.36  $ &$3.741\times 10^{-14}$ & $38.5005$ &  36.9018 & 35.8393 \\

NGC 4382 &$  1.30\times 10^{7}  $&$ 17.88 $  &$3.468\times 10^{-14}$& $34.7574 $ &  33.3141 & 32.3549 \\

        \end{tabular}
    \end{ruledtabular}
\end{table*}
\endgroup

\section{WEAK GRAVITATIONAL LENSING}
\begin{figure}[H]
    \begin{center}
                                \includegraphics[scale=0.7]{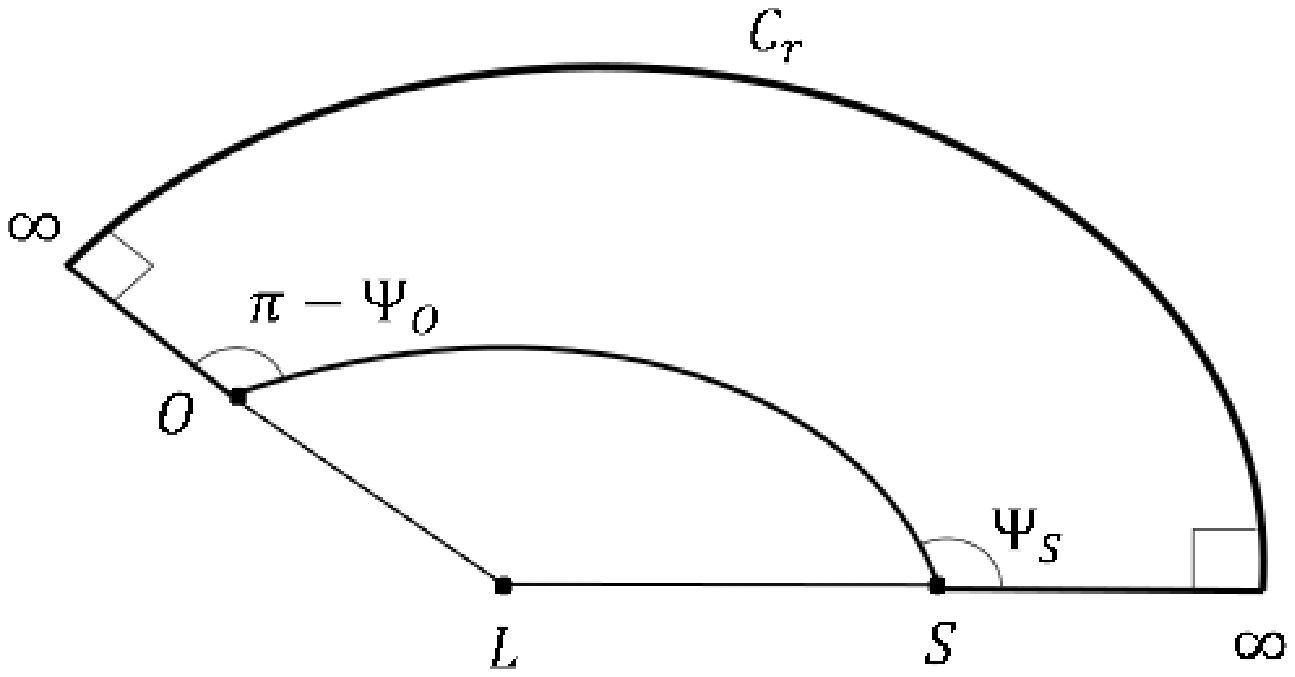}
        \caption{Schematic diagram of the quadilateral ${}_O^{\infty}\Box_{S}^{\infty}$. }\label{lensing}
    \end{center}
\end{figure}

Along with the strong lensing we should also study the weak
lensing scenarios. In this regard this section is devoted to study
the weak lansing scenario for this hairy black hole. Thus we
deduce the deflection angle in the weak field limit using the
Gauss-Bonnet theorem. The method was first proposed by Gibbons and
  Werner \cite{GW}. We assume that the black hole(L)
   is at the center of the coordinate system and the
   receiver( R) and the source (S) are at finite distances
   from the black hole.  Following the earlier composition
   \cite{ISHIHARA1, CARMO} and the assumption employed there, the
   deflection angle can written be written down as
\begin{equation}
\alpha_D=\Psi_R-\Psi_S+\Phi_{OS},
\end{equation}
where $\Phi_{OS}=\Phi_{O}-\Phi_{S}$, $\Psi_R$ is the angle made by
light rays at the receiver and $\Psi_S$ is the angle that is made
by light rays at the source. Now, we consider the quadrilateral
${}_O^{\infty}\Box_{S}^{\infty}$. It consists of spatial light ray
curves from  the source (S) to the observer (O), a circular arc
segment $C_r$  and two outgoing radial lines from the observer $O$
and the source $S$.  Vide the Fig. \ref{lensing}.
 Consequently the angle of deflection is given by
\begin{equation}
\alpha_D=-\int\int_{{}_R^{\infty}\Box_{S}^{\infty}} K dS.\label{deflectionangle}
\end{equation}
where K is the Gaussian curvature. For null geodesics
 we have $ds^2=0$ and thus, from Eqn. (\ref{metric}) we have
\begin{equation}
dt= \pm\sqrt{\gamma_{ij}dx^i dx^j},
\end{equation}
with
\begin{equation}
\gamma_{ij}dx^i dx^j=\frac{1}{f(r)^2}dr^2+\frac{r^2}{f(r)}\left(d\theta^2+\sin^2\theta\,
d\phi^2\right). \label{metric3}
\end{equation}

We now need the expression od Gaussian curvature of the surface
where light propagation which is is defined by \cite{WERNER}
\begin{eqnarray}
K&=&\frac{{}^{3}R_{r\phi r\phi}}{\gamma},\nonumber\\
&=&\frac{1}{\sqrt{\gamma}}
\left(\frac{\partial}{\partial \phi}
\left(\frac{\sqrt{\gamma}}{\gamma_{rr}}{}^{(3)}\Gamma^{\phi}_{rr}\right)
 - \frac{\partial}{\partial r}\left(\frac{\sqrt{\gamma}}{\gamma_{rr}}{}^{(3)}
 \Gamma^{\phi}_{r\phi}\right)\right),\label{gaussian}
\end{eqnarray}
where $\gamma=\det(\gamma_{ij})$. For the
 hairy Schwarzschild black hole metric (\ref{metric}),
  in the weak-field limits,  Eqn. (\ref{gaussian}) in
  the leading order terms yields
\begin{equation}
K=\frac{3 M^2}{r^4}-\frac{2 M}{r^3}+\alpha \left(\frac{3\ell_{0}}{2 M r^2}-\frac{2 M}{r^3}
+\frac{3}{r^2}\right)+\mathcal{O}\left(\frac{\alpha^2}{M^2},\frac{\alpha^2\ell_0}{M^3},
\frac{\alpha^2\ell_0^2}{M^4}\right)
\end{equation}
The surface integral of Gaussian curvature over the closed
quadrilateral ${}_O^{\infty}\Box_{S}^{\infty}$ reads \cite{ONO1}
\begin{equation}
\int\int_{{}_O^{\infty}\Box_{S}^{\infty}} K dS =
\int_{\phi_S}^{\phi_O}\int_{\infty}^{r_0} K \sqrt{\gamma}dr
d\phi,\label{Gaussian}
\end{equation}
where $r_0$ is the distance of closest approach
to the black hole. If we assume that the light rays
follow straight line trajectory i.e $r=\frac{b}{sin\phi}$,
 then the deflection angle is given by
\begin{equation}
\alpha_D^0=\frac{4 M}{b}
+\alpha \left(\frac{15 \pi\ell_0 M}{16 b^2}+\frac{3\ell_0}{2 b}+\frac{M}{b}\right)
\end{equation}
where we have retained terms up to the order of M. With help of
the improved article \cite{CRISNEJO}, to include the higher order
 corrections, we take the trajectory
\begin{equation}
u=\frac{\sin\phi}{b} + \frac{M(1-\cos\phi)^2}{b^2}-\frac{M^2(60\phi\,
\cos\phi+3\sin3\phi-5\sin\phi)}{16b^3}+\mathcal{O}\left( \frac{M^2\alpha}{b^5}\right)
,\label{uorbit}
\end{equation}
where $u=1/r$, and $b$ is the impact parameter. The integral
Eqn.(\ref{Gaussian}) can be recast as
\begin{equation}
\int\int_{{}_O^{\infty}\Box_{S}^{\infty}} K dS =
\int_{0}^{\pi+\alpha_D^0}\int_{0}^{u}-\frac{K\sqrt{\gamma}}{u^2}du
d\phi,
\end{equation}
which results
\begin{eqnarray}\nonumber
\alpha_D&=&\frac{4 M}{b}+\frac{15 \pi  M^2}{4 b^2}
+\frac{128 M^3}{3 b^3}+\alpha \left(\frac{3 M^3 (1792 b+3985 \pi\ell_0)}{256 b^4}+\frac{M^2
(9 \pi  b+167 \ell_0)}{6 b^3}+\frac{M (16 b+33 \pi \ell_0)}{16 b^2}+\frac{3\ell_0}{2 b}\right)\\
&&+\mathcal{O}\left(\frac{M^4}{b^4},\frac{M^4 \alpha}{b^4},\frac{M^4 \alpha \ell_0}{b^5} \right)
\end{eqnarray}
Note that in the limit $\alpha=0$, the above equation reduces to
the value for the Schwarzschild black hole \cite{VIRBHADRA,
CRISNEJO}
\begin{equation}
\left.\alpha_D\right|_{\text{sch}}=\frac{4M}{b}+\frac{15\pi M^2}{4b^2}
+\frac{128M^3}{3b^3}+\mathcal{O}\left(\frac{M^4}{b^4} \right),
\end{equation}
before ending our discussion we give of graphical presentatio of
variation of deflection in the weak gravitational limit with the
impact parameter.
\begin{figure}[H]
    \begin{center}
        \begin{tabular}{c c}
            \includegraphics[scale=0.7]{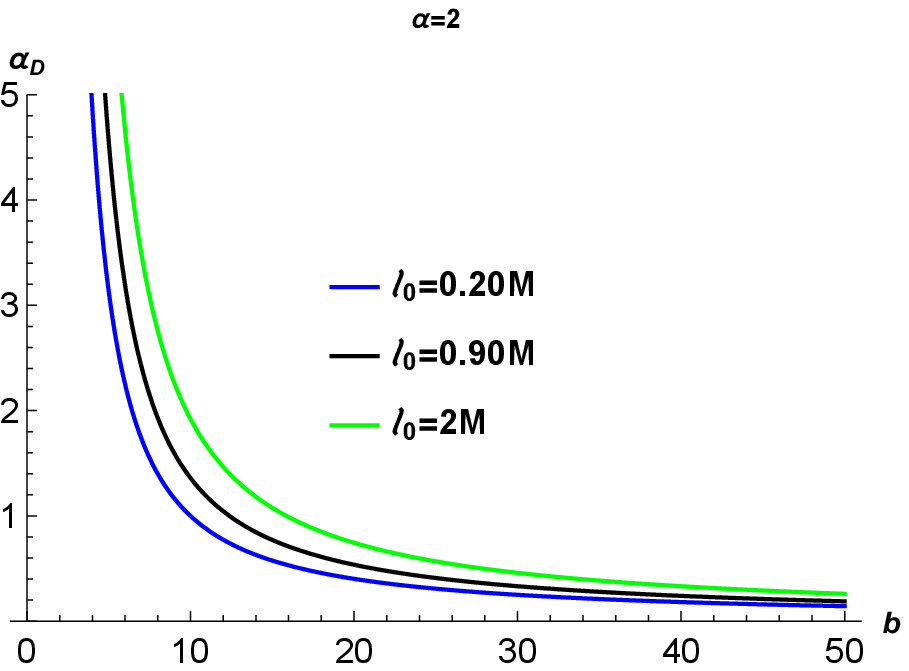}&
            \includegraphics[scale=0.7]{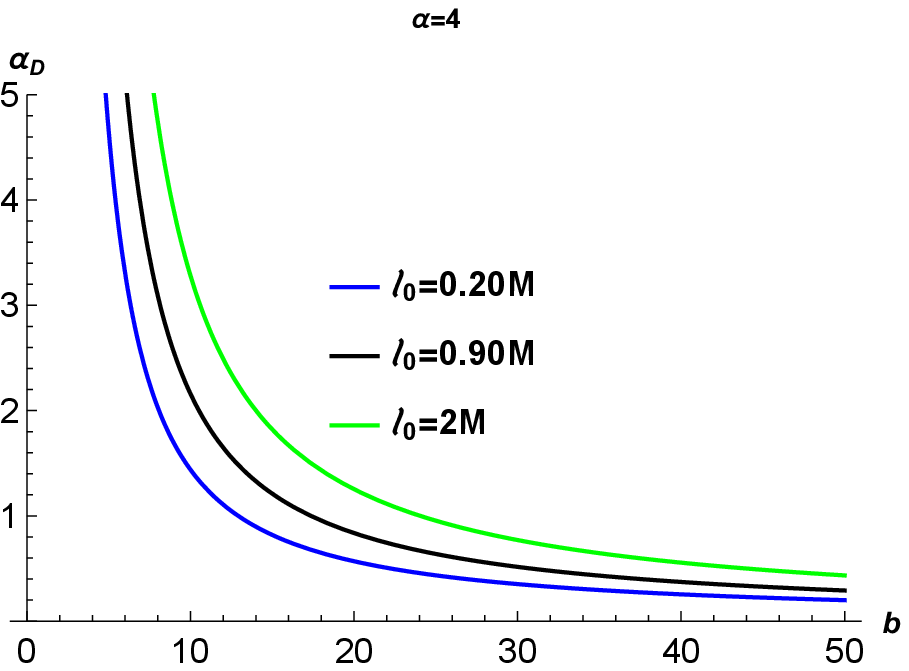}\\
        \end{tabular}
        \caption{Deflection angle for weak gravitational lensing. The
        Left one is for $\alpha=2$ and the right one is for $\alpha=4$.
        Here we have taken $M=1$. }\label{weak}
    \end{center}
\end{figure}
The plots transpire that the deflection angle in weak
gravitational lensing decreases with the impact parameter $b$ but
increases with $\ell_0$ and $\alpha$
\section{summary and conclusion}
Black holes have been considered more than bare fantastic results
of the Einstein equations for quite some time now. Nonetheless, it
is only recently that their direct existence was detected,
substantially due to the report of the results of the Event
Horizon Telescope Collaborations. So the study of aspects of black
holes and the consequences of their presence have now surfaced
with new alleviation. In this environment study of strong as well
as weak lensing in the background of a modified black hole is of
interest since it is a realistic picture to consider the griding
of a black hole is not a vacuum rather it is encircled by matter
field of various kind. Hence mollification due to scalar hair is
considered here and an attempt has been made to study lensing
scenarios and how the presence of hair influence the lensing
property has been studied in detail. The rationale adopted here
evades the no-hair theorem since the hair conceived in this
situation by fresh source from the surrounding, such as dark
matter, which is considered to have has constant energy-momentum
tensor. This modification permitted us two unfastened parameters
$\alpha$ and $\ell_0$. The former refers back to the deformation
because of the presence of outside supply and the after bone is
associated with alteration in thermodynamical parcels because of
the presence of hair.

We observe that the lensing coefficient $\overline{a}$ increases
with $\alpha$ but lensing coefficient $\overline{b}$ and the
impact parameter $b_m$ decrease with the increase of $\alpha$.
However, with the increase of  $\ell_0$, although  $\overline{a}$
decrease $\overline{b}$ gets enhanced. The impact parameter $b_m$
increases with the increase of $\ell_0$. The values of the
parameters resembles Schwarzschild black hole whenever we put
$\alpha=0$. Our investigation indicates a conspicuous change in
the lensing properties due to the presence of the hair which is
indeed for the non-rotation black hole. We admit that the rotating
black hole is extra near to the truth than that of the
non-rotating one. Still, the perceptible changes which have been
determined here can now no longer be ignored.


\begin{thebibliography}{the}
\bibitem{EINST}  A. Einstein, Science, 84 (1936) 506.
\bibitem{EHT1} K. Akiyama et al. (Event Horizon Telescope Collaboration),
First M87 Event Horizon Telescope results. I. The shadow of the
supermassive black hole, Astrophys. J. 875, L1 (2019).
\bibitem{EHT2} K. Akiyama et al. (Event Horizon Telescope
Collaboration), First M87 Event Horizon Telescope results. II.
Array and instrumentation, Astrophys. J. 875, L2 (2019).
\bibitem{EHT3} K.
Akiyama et al. (Event Horizon Telescope Collaboration), First M87
Event Horizon Telescope results. III. Data processing and
calibration, Astrophys. J. 875, L3 (2019).
\bibitem{EHT4} K. Akiyama et al.
(Event Horizon Telescope Collaboration), First M87 Event Horizon
Telescope results. IV. Imaging the central supermassive black
hole, Astrophys. J. 875, L4 (2019).
\bibitem{EHT5}K. Akiyama et
al. (Event Horizon Telescope Collaboration), First M87 Event
Horizon Telescope results. V. Physical origin of the asymmetric
ring, Astrophys. J. 875, L5 (2019).
\bibitem{EHT6} K. Akiyama et
al. (Event Horizon Telescope Collaboration), First M87 Event
Horizon Telescope results. VI. The shadow and mass of the central
black hole, Astrophys. J. 875, L6 (2019)
\bibitem{VERG} S.U. Viergutz, A. A. 272 (1993) 355.
 \bibitem{BAR} J.M. Bardeen, Black Holes, ed. C. de Witt  B.S. de Witt, NY, Gordon  Breach, 215 (1973).
\bibitem{FLAKE} H. Falcke, F. Melia, E. Agol, ApJ Letters 528 (1999) L13.
\bibitem{VRIB} K.S. Virbhadra, G.F.R. Ellis, Phys. Rev. D62 (2000) 084003.
 \bibitem{FRIT} S. Frittelli, T.P. Kling, E.T.Newman, Phys. Rev. D 61 (2000) 064021
\bibitem{BOZZ} V. Bozza, S. Capozziello, G. Iova,  G. Scarpetta, Gen. Rel. and Grav. 33 (2001) 1535.
\bibitem{EIRO} E.F. Eiroa, G.E. Romero, D.F. Torres,
\bibitem{ELL} K.S. Virbhadra, G.F.R. Ellis, Phys. Rev. D65 (2002) 103004.
\bibitem{GIBB} G. W. Gibbons, M. C. Werner: Class. Quant. Grav., 25:235009, 2008.
\bibitem{AGV} A. Grenzebach, V. Perlick,  C. Lammerzahl: Phys. Rev., D89(12):124004, 2014.
\bibitem{CMW} M. C. Werner: Gen. Rel. Grav., 44:3047-3057, 2012.
\bibitem{ISHIHARA1} A. Ishihara, Y. Suzuki, T. Ono, T. Kitamura,  H. Asada. Phys. Rev., D94(8):084015, 2016.
\bibitem{ISHIHARA2}A. Ishihara, Y. Suzuki, T. Ono and H. Asada, Phys. Rev. D 95, 044017 (2017).
\bibitem{ONO1}T. Ono, A. Ishihara and H. Asada, Phys. Rev. D 96, 104037 (2017).
\bibitem{ONO2} T. Ono, A. Ishihara, H. Asada:  Phys. Rev., D96(10):104037, 2017.
\bibitem{ONO3} T. Ono, A. Ishihara,  H. Asada. Phys. Rev. D98(4):044047, 2018.
\bibitem{OVALLE1}J. Ovalle, R. Casadio, E. Contreras, and A. Sotomayor, Phys. Dark Univ. 31, 100744 (2021).
\bibitem{OVALLE2}J. Ovalle, Phys. Rev. D 95, 104019 (2017).
\bibitem{OVALLE3}J. Ovalle, Phys. Lett. B 788, 213 (2019); J. Ovalle, Mod. Phys. Lett. A 23, 3247 (2008).
\bibitem{OVALLE4}E. Contreras, J. Ovalle, and R. Casadio, Phys. Rev. D 103, 044020 (2021).
\bibitem{BOZZA}V. Bozza, Phys. Rev. D 66, 103001 (2002).
\bibitem{BOZZA1}V. Bozza, S. Capozziello, G. Iovane and G. Scarpetta, Gen. Rel. Grav. 33, 1535 (2001).
\bibitem{BOZZA2}V. Bozza and L. Mancini, Gen. Rel. Grav. 36, 435 (2004).
\bibitem{GW} G. W. Gibbons and M. C. Werner, Class. Quant. Grav. 25, 235009 (2008).
\bibitem{CHANDRA}S. Chandrasekhar, The Mathematical Theory of Black Holes (Oxford University Press, New York, 1992).
\bibitem{VIRBHADRA}K. S. Virbhadra and G. F. R. Ellis, Phys. Rev. D 62, 084003 (2000).
\bibitem{WEINBERG}S. Weinberg, Gravitation and Cosmology: Principles and Applications of the General Theory of Relativity (New York:Wiley, 1972).
\bibitem{GHOSH}S. G. Ghosh, R. Kumar, and S. U. Islam, J.Cosmol. Astropart. Phys. 03, 030 (2021).
\bibitem{CARMO}M. P. Do Carmo, Differential Geometry of Curves and Surfaces, (Prentice-Hall, New Jersey, 1976).
\bibitem{WERNER} M. C. Werner, Gen. Rel. Grav. 44, 3047 (2012).
\bibitem{TD} T. Do et al., Science 365, 664 (2019).
\bibitem{EINSTEIN}A. Einstein, Science, 84, 506 (1936).
\bibitem{CRISNEJO} G. Crisnejo, E. Gallo and K. Jusufi, Phys. Rev. D 100,104045 (2019).
\end{thebibliography}
\end{document}